# Why so many sperm cells?

K. Reynaud[1]  Z. Schuss[2], N. Rouach[3], D. Holcman[4]

[1]Biologie du Développement et Reproduction, Ecole Nationale Vétérinaire d'Alfort, 7 avenue du Général de Gaulle 94704 Maisons-Alfort Cedex

[2]Department of Applied Mathematics, Tel-Aviv University, 69978 Tel-Aviv, Israel.

[3]Center for Interdisciplinary Research in Biology, INSERM U1050, CNRS UMR 7241, Collège de France, 75005 Paris, France.

[4]Group of Computational Biology and Applied Mathematics, Institute for Biology, Ecole Normale Supérieure, 75005 Paris, France.

**Abstract:**

A key limiting step in fertility is the search for the oocyte by spermatozoa. Initially, there are tens of millions of sperm cells, but a single one will make it to the oocyte. This may be one of the most severe selection processes designed by evolution, whose role is yet to be understood. Why is it that such a huge redundancy is required and what does that mean for the search process? we propose to discuss here these questions and consequently a new line of interdisciplinary research needed to find possible answers.

**Main text:**

A key limiting step in fertility is the search for the oocyte by spermatozoa. Initially, tens of millions of sperm cells depart from the vagina to the uterus, but a single one will make it to the oocyte. This may be one of the most severe selection processes designed by evolution, whose role is yet to be understood. Why is it that such a huge redundancy is required and what does that mean for the search process? These questions are more relevant now, since recent studies [1] have demonstrated that, between 1989 and 2005, the concentration of sperm cells in human semen has significantly and continuously dropped at an average rate of 1,9% a year, leading to a reduction of 32,2% in sperm count over 16 years. Some other sperm characteristics appear to have suffered as well, such as sperm motility or the percentage of morphologically normal spermatozoa. Does the decline in sperm count really matter? Some framework appears desirable to properly evaluate the effect of this decline.

We suggest that these questions may be narrowed down to that of a search for a small target (the oocyte) by random motion of the sperm cells. The mean search time $\tau$ can be computed. For Brownian motion, the narrow escape theory [2] gives a first order approximation

$$\tau = V/4Da,$$

where $V$ is the volume of the uterus, $D$ is the effective diffusion constant of sperm cells and $a$ is the radius of the oocyte Indeed, it is only close to the oocyte (within a few tenths of micrometer) that a chemotaxis signal may be expected to affect the sperm cells. Some consideration has also to be given to the complex geometrical shape of the uterus, to the location of the oocyte far down the fallopian tube and to the limited access from the uterus to the oviduct through the utero-tubal junction.

The purpose of starting the search with a large number $N$ of independent spermatozoa seems to be the decrease in the mean search time $\tau$ to $\tau/N$. In other biological systems, finding a small hidden target by random search is often achieved by a large redundancy in the number of searchers. This is the case, for instance, with identical neurotransmitters released from a vesicle into the cleft of a neuronal synapse. The vesicles released from the presynaptic terminal contain $2\text{-}4\times10^3$ neurotransmitter molecules, which have to find, by Brownian motion, receptors (small targets) on the surface of the postsynaptic terminal at a distance of 20 nm away from the release site. There are on average less than 50 receptors, such as glutamate receptors, on an excitatory postsynaptic terminal, which is disproportionately smaller than the number of released neurotransmitters. The excess in released neurotransmitters guarantees that a certain number of receptors will eventually bind to neurotransmitters and thus open for the transmission of ions, leading to a synaptic response with a probability close to one in a timescale of less than a millisecond. Indeed, the probability for a Brownian neurotransmitter, released exactly at the center of the presynaptic terminal, to find a receptor is in the range of $10^{-4}$ to $10^{-3}$. Interestingly, the number of neurotransmitter molecules in the vesicle is in the order of 3000, which exactly compensates for the low probability. Ultimately, the number of open receptors is in the order of a few. Redundancy seems to be the more adequate response when a target has to be found by random searchers. The exact number seems to depend on the probability to find the target before a certain time.

Using a scaling law argument, it is tempting to think that the process leading a sperm cell to find the oocyte may be somewhat similar to the process leading a neurotransmitter molecule to find a receptor or any search process by random searchers. The number of imaginary searcher that would be necessary to find a small target in a space the size of the uterus would be $3\times10^8$ bigger than that in a synapse. We obtain a range in the same order of magnitude as the number of sperm cells usually found in a human ejaculate (ca 280 millions) for a uterus of 40 cm$^3$ volume. This rough estimate based on the Brownian motion of sperm gives a correct order of magnitude, suggesting that the uterus might create the condition of sufficient randomness. Thus the initial number of deposit spermatozoa might ignore a fundamental inherent difference that sperm cells are not equipotent in terms of fertilizing potential. Sperm cells are genetically different from each other. Some of those differences may affect their motility and/or their affinity for the oocyte (or their response to chemotactic attraction by the COC). These differences should be further included to study the consequences on the efficacy of fertilization.

The sperm motion, however, is not Brownian. Actually, the above considerations raise several questions. First, the motion of sperm cells in the uterus has to be tracked *in vivo* by new microscopy techniques that can reveal not only the type of spermatozoidal trajectory, but also its position in the uterus and the structure of the endometrium. Actually, sperm cells may not fill the uterine lumen or the various parts of the female genital tract, but could rather be concentrated near the surface in a superficial layer, as observed in microchambers [3]. Second, the uterus is far from being a homogenous environment and it is not clear how the structure of the endometrium affects the motion of sperm cells. Breaking the limit in resolution of current optical devices will certainly be a key step in reconstructing sperm cell trajectories in utero.

Other factors should also be accounted for such as the fluid micro-environment generated by secretions in the cervix, uterus and oviduct. The relationship between these different factors still requires clarification. The connections if any between the dynamical properties such as viscosity, cell local motion, muscular contractions in the female genital tract and the biochemical factors such as pH, chemotaxis, fluid composition, endocrine factors (stage in the cycle, hormones and growth factors , systemic vs. local paracrine/endocrine farctors), and immunological factors are yet to be found. For example, are epithelial cells capable of identifying key characteristics of gamete cells? Understanding the role of these factors will pave the way for clarifying the tedious and long journey of the spermatozoa to the oviduct characterized by constant screening in the cervix through the uterus and at the utero-tubal junction. Interestingly, during their transit from the vagina to the oviduct, the spermatozoa number is drastically reduced by a factor of roughly 1000, an estimate that should be reanalyzed by incorporating the factors mentioned above.

In this entire process, what seems to really matter is the number of sperm cells that reach the immediate vicinity of the oocyte which depends on the number of sperm cells injected into the vagina. By integrating all physiological factors and dynamical information extracted from single particle trajectories, theoretical modeling should contribute to estimate the continuous loss of sperm cells between the release of sperm into the vagina and the immediate environment of the cumulus oocyte complex (COC) in the fallopian tube. It might be conceivable that sperm cell motility may have little to do with the success of passing through the cervical canal and the rapid transit though the uterus. Aspiration by post-coital contractions of muscular fibers in the uterine wall may account for most of the rapid displacement of sperm to the utero-tubal junction. Such possibility might have to be examined as well. Actually, to start with, a significant amount of sperm is lost when the penis is withdrawn from the vagina, especially if the penis is still erect when withdrawn, a regrettably common practice nowadays. Indeed, a morphological difference between humans and rats is that woman does not have a vaginal plug to keep sperm cells in. Some sperm remains on the walls of the vagina in any case and, among the sperm cells that make it to the cervical canal, only a fraction, to be estimated, reach the uterine cavity. Finally, close the ovule, spermatozoa motion is certainly reprogrammed but little is known about possible exchanges of chemical information between the oocyte and the spermatozoa that leads to the ultimate spermatozoa selection.

All these constructive considerations cannot explain why a drop in sperm count from 100 million to 20 million per ml leads to infertility. Explaining the problem of search for oocytes will require a combination of approaches, starting with physiological data, mathematical modeling analysis of spermatozoidal trajectories and more advanced observation techniques. This viewpoint is designed to attract the attention of the scientific community to specific aspects of the search process for oocytes, which is a fundamental step in animal reproduction. Whatever progress can be made in understanding this search process should lead to a better understanding of the causes of infertility.